\documentclass{article}

\usepackage{PRIMEarxiv}

\usepackage[utf8]{inputenc} 
\usepackage[T1]{fontenc}    
\usepackage{hyperref}       
\usepackage{url}            
\usepackage{booktabs}       
\usepackage{amsfonts}       
\usepackage{nicefrac}       
\usepackage{microtype}      
\usepackage{lipsum}
\usepackage{fancyhdr}       
\usepackage{graphicx}       

\usepackage{epstopdf}
\usepackage[caption=false]{subfig}

\usepackage{natbib}
\bibpunct[, ]{(}{)}{;}{a}{}{,}

\usepackage[utf8]{inputenc} 
\usepackage[T1]{fontenc}    
\usepackage{hyperref}       
\usepackage{url}            
\usepackage{booktabs}       
\usepackage{amsfonts}       
\usepackage{amsmath}
\usepackage{amsthm}
\usepackage{bm}
\usepackage{nicefrac}       
\usepackage{lipsum}		
\usepackage{graphicx}
\usepackage{natbib}
\usepackage{doi}
\usepackage{subcaption}
\usepackage{adjustbox}
\usepackage[dvipsnames]{xcolor}

\usepackage{algorithm} 
\usepackage{algpseudocode} 

\newcommand{\proglang}[1]{\textbf{\texttt{#1}}}
\newcommand{\pkg}[1]{\textbf{#1}}
\newcommand{\code}[1]{\texttt{#1}}

\theoremstyle{plain}

\theoremstyle{definition}

\theoremstyle{remark}

\DeclareMathOperator{\argmin}{argmin}

\graphicspath{{media/}}     

\title{Cellwise robust and sparse principal component analysis
}

\author{
  Pia~Pfeiffer, Laura~Vana-Gür,  and Peter~Filzmoser \\
  Institute of Statistics and Mathematical Methods in Economics \\
  TU Wien \\
  Vienna, Austria\\
  \texttt{pia.pfeiffer[at]tuwien.ac.at} \\
}

\begin{document}
\maketitle

\begin{abstract}
A first proposal of a sparse and cellwise robust PCA method is presented. Robustness to single outlying cells in the data matrix is achieved by substituting the squared loss function for the approximation error by a robust version. The integration of a sparsity-inducing $L_1$ or elastic net penalty offers additional modeling flexibility.
For the resulting challenging optimization problem, an algorithm based on Riemannian stochastic gradient descent is developed, with the advantage of being scalable to high-dimensional data, both in terms of many variables as well as observations. The resulting method is called SCRAMBLE (Sparse Cellwise Robust Algorithm for Manifold-based Learning and Estimation).
Simulations reveal the superiority of this approach in comparison to established methods, both in the casewise and cellwise robustness paradigms. 
Two applications from the field of tribology underline the advantages of a cellwise robust and sparse PCA method.
\end{abstract}


\section{Introduction}

The increasing prevalence of large data sets, especially high-dimensional data in the sense of many more variables than observations, motivates the use and development of dimension-reduction techniques. Principal Component Analysis (PCA), dating back to \citet{Pearson1901} and \citet{Hotelling1933}, is one of the oldest and most widely applied dimension-reduction techniques. 
The idea of PCA is to find a low-dimensional representation of the data set in a way that preserves as much variance as possible~\citep[e.g.][]{Jolliffe2003}. 

Based on a mean-centered (and possibly scaled) data matrix $\bm X$, with $n$ observations in the rows, and $p$ variables in the columns, the principal components (PCs) are defined by the linear combination
$\bm Z=\bm X \bm V$ under the constraint that the columns of the $p\times p$ matrix $\bm V$ are normed to length 1 and orthogonal to each other.
Since the variances of the columns of $\bm Z$ have to be maximized,
the solution for $\bm V$ can be obtained by the spectral composition $\hat{\bm \Sigma}=\hat{\bm V} \hat{\bm A} \hat{\bm V}'$,
where $\hat{\bm \Sigma}$ is the estimated covariance matrix  of 
$\bm X$, and $\hat{\bm A}=\mbox{Diag}(\hat{a}_1,\ldots,\hat{a}_p)$ is the diagonal matrix with the corresponding estimated eigenvalues, arranged in descending order~\citep{Jolliffe2003}. 
The matrix $\hat{\bm V}$ is also known as loadings matrix, 
while $\hat{\bm Z}=\bm X \hat{\bm V}$ refers to the PCA score matrix.
Traditionally, the sample covariance matrix 
$\bm S = \frac{1}{n-1} \bm X' \bm X$ is used for $\hat{\bm \Sigma}$,
resulting in eigenvectors $\tilde{\bm V}$, eigenvalues
$\tilde{\bm A}$, and classical principal components
$\tilde{\bm Z}=\bm X \tilde{\bm V}$. The identical solution 
$\tilde{\bm V}$ can be obtained from a singular value decomposition (SVD) of
$\bm X$ as $\bm X=\tilde{\bm U} \tilde{\bm D} \tilde{\bm V}'$~\citep[e.g.][]{Jolliffe2003}.

As the main interest is usually in the first $k$ PCs, where
$k<p$, or even $k\ll p$ for high-dimensional data, it is not necessary to compute the whole $p\times p$ matrix $\hat{\bm V}$,
but to only focus on the matrix 
$\hat{\bm V}_k \in \mathbb{R}^{p\times k}$ with the first $k$ columns,
to obtain the first $k$ PCs $\hat{\bm Z}_k=\bm X \hat{\bm V}_k$.
Especially for $p\gg n$, the approach based on a spectral decomposition of the estimated covariance matrix is numerically 
not attractive, and thus SVD is commonly employed in this case.
This leads to a rank-$k$ approximation $\tilde{\bm X}_k = \tilde{\bm U}_k \tilde{\bm D}_k \tilde{\bm V}_k'$ of $\bm X$, 
with $\tilde{\bm U}_k \in \mathbb{R}^{n\times k}$,
$\tilde{\bm V}_k \in \mathbb{R}^{k\times k}$
and $\tilde{\bm D}_k \in \mathbb{R}^{k\times k}$ with elements $\tilde{d}_{ii} \geq 0$ for $i = 1, \ldots, k$ and $\tilde{d}_{ij} = 0$ otherwise.
The rank-$k$ SVD is the best rank-$k$ approximation in the Frobenius norm \citep{Eckart1936}, 
\begin{align}
\tilde{\bm V}_k = \argmin_{\bm V_k} \| \bm X - \bm X \bm V_k \bm V_k' \|^2_F \label{eq:SVD_RSS}
\end{align}
for any $p\times k$ matrix $\bm V_k$ with $\mbox{rank}(\bm V_k) \leq k$ and $\bm V_k' \bm V_k =\bm I_k$.
We can define the residual matrix
\begin{align}
\label{eq:residuals}
\bm R=\bm X - \bm X \bm V_k \bm V_k' \quad 
\mbox{with elements}
\quad r_{ij}, \mbox{ for } i=1,\ldots ,n \mbox{ and } j=1,\ldots ,p, 
\end{align}
and problem~\eqref{eq:SVD_RSS} is equivalent to minimizing 
the sum of all squared residuals by using a matrix $\bm V_k$ as defined above. 

It is well known that a
sum-of-squares criterion is sensitive to outlying entries~\citep{Maronna2019}.
Such outliers, however, refer to single outlying cells
in the residual matrix $\bm R$, and not necessarily to outlying rows (observations). The latter is traditionally considered for robustly estimating a covariance matrix~\citep[e.g.][]{Maronna1976,Rousseeuw1985,Rousseeuw2005} in order to obtain robust PCs with the help of spectral decomposition~\citep{Maronna2019}, but in the case $p>n$ those estimators cannot be computed.

The concept of cellwise outliers has been introduced in~\cite{Alqallaf2009}, but already~\cite{Maronna2008} formulated a cellwise robust PCA version by downweighting outlying cells in the residual matrix rather than outlying rows. 
In fact, they proposed to replace the Frobenius norm in~\eqref{eq:SVD_RSS} with a more robust loss function.
Cellwise robustness has particular advantages if $p\gg n$: Assuming that any data cell has the same chance to be contaminated, even a small amount of contamination can lead to many rowwise outliers, which at some point could even form the majority and cause breakdown of traditional rowwise robust methods~\citep{Maronna2019}.
A robust PCA method that cannot only deal with cellwise outliers but also with missing values is MacroPCA \citep{Hubert2019}.
This method combines the DetectDeviatingCells (DDC) algorithm of~\citet{Rousseeuw2018} to detect cellwise outliers with a version of ROBPCA~\citep{Hubert2005}, a robust PCA method that can also deal with high-dimensional data. 

None of the proposed cellwise robust PCA methods lead to sparse solutions. The contribution of this paper is sparsity: In the high-dimensional case it is desirable to obtain (many) zeros in the matrix $\bm V_k$, since this simplifies the interpretation of the PCs. In this paper we will introduce a cellwise robust and sparse PCA method.
We first review robust and rowwise sparse PCA methods in Section~\ref{sec:related} before introducing our method in Section~\ref{sec:method}. An algorithm for its computation based on manifold learning is presented in Section~\ref{sec:algo},  Section~\ref{sec:sim} provides a comparison of simulation results with alternative PCA methods. Applications in Section~\ref{sec:exa} demonstrate the usefulness of the method. The final Section~\ref{sec:sum} provides a summary and conclusions.

\section{Related work}
\label{sec:related}

Robust PCA has been a very active research field, and thus many different approaches are available in the literature~\citep[see, e.g.,][for an overview]{She2016}. Since our focus is also on sparsity, we provide a short and thus possibly incomplete overview of robust and sparse PCA methods in the following, before discussing cellwise robust methods in this context.

One of the first papers on robust sparse PCA is~\citet{Candes2011}. However, sparsity here refers to a sparse residual matrix as the remainder of a robust rank-$k$ approximation. In this paper, we are more interested in sparsity for the loadings matrix, as this simplifies the interpretation of the PCs. Although this work is highly visible, the method is intended to deal with additive outliers, and not with outliers in the orthogonal complement to the PCA subspace, see~\citet{She2016,Hubert2005}.  

As PCA can also be seen as a projection-pursuit (PP) problem, with the task to search for a projection direction that maximizes the variance of the projected observations, \citet{Croux2005} introduced a robust procedure by considering a robust variance (scale) estimator. Inspired by the
SCoTLASS approach which adds a LASSO penalty on the direction vectors~\citep{Jolliffe2003}, this PP approach was also reformulated to include an $L_1$ penalty, resulting in a robust and sparse PCA method~\citep{Croux2013}.

The ROBPCA method by \citet{Hubert2005} combines a PP approach with the plug-in method by first projecting the data to a low-dimensional subspace and then applying a robust estimator of the covariance. 
This method was extended in~\citet{Hubert2016}, who proposed ROSPCA
by integrating an $L_1$-penalty with the ROBPCA algorithm, which also leads to sparse solutions. 

\citet{Zou2006} suggest reformulating PCA as a regression problem. Then, sparse loadings can be derived using elastic net \citep{Zou2003} and LASSO \citep{Tibshirani1996} regression.
A robust extension that is based on robust plug-in estimators for the covariance was proposed by \citet{Greco2016}. In the case $p > n$ they propose to use the unconstrained ROBPCA solution to obtain a plug-in estimator.

\section{Cellwise robust sparse PCA for high-dimensional data}
\label{sec:method}

Following the ideas outlined in the previous section, the rank-$k$ approximation criterion~\eqref{eq:SVD_RSS} can be combined with a criterion to obtain sparsity. 
When using an elastic net penalty, the modified problem formulation is
\begin{align}
    \hat{\bm V}_k = \argmin_{\bm V_k' \bm V_k = \bm I_k} \| \bm X - \bm X \bm V_k \bm V_k' \|^2_F + \sum_{j = 1}^k \lambda_j (\alpha \| \bm v_j \|_2^2 + (1 - \alpha) \| \bm v_j \|_1), \label{eq:SVD_RSS_sparse}
\end{align}
where $\bm v_j$ refers to the $j$-th column of $\bm V_k$, $1\leq j \leq k< p$, the strength of regularization for each component is controlled by $\bm \lambda = (\lambda_1, \ldots, \lambda_k)'$, and the elastic net mixing parameter $\alpha$ controls the sparsity~\citep{Zou2005}. 

As the least squares loss is highly susceptible to outlying observations, we propose to substitute it with a robust loss function, similar to the suggestion in~\citet{Maronna2008}, in order to obtain a cellwise robust and sparse PCA method:
\begin{align}
    \hat{\bm V}_k = \argmin_{\bm V_k' \bm V_k = \bm I_k} \frac{1}{np}\sum_{j = 1}^p \hat{\sigma}_j^2
    \sum_{i = 1}^n\rho \left (\frac{r_{ij}}{\hat{\sigma}_j} \right ) + \sum_{j = 1}^k \lambda_j (\alpha \| \bm v_j \|_2^2 + (1 - \alpha) \| \bm v_j \|_1), \label{eq:SVD_RSS_sparse_robust}
\end{align}
where $r_{ij}$ are the residuals from~\eqref{eq:residuals}, 
and $\hat{\sigma}_j$ 
is a column-wise estimator of residual scale. 
The function $\rho$ corresponds to a robust loss function~\citep{Maronna2019}, and popular choices are the Huber loss, defined as 
\begin{align}
    \rho_H(r) = \begin{cases} 
         r^2 &\text{for~} |r|\leq b,\\
         b |r| &\text{otherwise},
       \end{cases} \label{eq:Huber}
\end{align}
the Tukey loss, defined as
\begin{align}
    \rho_T(r) = \begin{cases} 
         \left (\frac{r}{c}\right )^2 \left (3- 3\left (\frac{r}{c}\right)^2 + \left (\frac{r}{c}\right)^4\right ) &\text{for~} |r|\leq c,\\
         1 &\text{otherwise},
       \end{cases} \label{eq:Tukey}
\end{align}
or a trimmed version of the least squares loss, defined as
\begin{align}
    \rho_{LTS}(r) = \begin{cases} 
         r^2 &\text{for~} |r|\leq |r|_{(h)},\\
         0 &\text{otherwise},
       \end{cases} \label{eq:LTS}
\end{align}
where $|r|_{(i)}$ refers to the ordered values of the absolute residuals, i.e.,~$|r|_{(1)} \leq \ldots \leq |r|_{(n)}$ and $h \in \lfloor \nicefrac{n}{2}, n \rfloor$. The trimming is applied column-wise. 

The choice of the loss function will also determine the robustness properties of $\hat{\bm V}_k$, see also~\citet{Maronna2008} for more detailed discussions.
For either of these loss functions, the influence of data points with large scaled residuals is reduced, resulting in a more robust estimate~\citep{Maronna2019}.
Appropriate parameter choices for the constants $b$ and $c$ in \eqref{eq:Huber}
and \eqref{eq:Tukey}, respectively, are proposed in the literature~\citep{Maronna2019}.
Throughout our experiments, we will use the default parameters 
$b=1.35$, $c=1.35$, assuming that the regular observations are standard Gaussian and $h=0.5$, {leading to maximum robustness for trimming.

Note that a multiplication with $\hat{\sigma}_j^2$
in the objective function~\eqref{eq:SVD_RSS_sparse_robust} is
important, since this guarantees that for the special choice 
$\rho(r)=r^2$ one obtains the objective function~\eqref{eq:SVD_RSS_sparse}
of the non-robust version. The estimation of the residual scale will be discussed in detail in the next section.

For outlier diagnostics, i.e.,~for distinguishing between regular observations, good and bad leverage points, and orthogonal outliers, a diagnostic plot can be constructed as proposed by \citet{Hubert2005}. The score distances (SD, i.e.,~Mahalanobis-like measure of
distance of an observation within the PC space) and orthogonal distances (OD, i.e.,~the orthogonal distance of an observation to the space spanned by the first $k$ PCs) are computed based on the robust PCA result and plotted together with the cutoff values, also described in detail in \citet{Hubert2005}.

\section{Algorithm}
\label{sec:algo}

Problem \eqref{eq:SVD_RSS_sparse_robust} is an optimization problem under constraints.
The objective function 
\begin{align}
    \mathcal{L}(\bm V) = \frac{1}{np} \sum_{j=1}^p
    \hat{\sigma}_j^2 \sum_{i=1}^n
    \rho\left(\frac{r_{ij}(\bm V)}{\hat{\sigma}_j}\right) + \sum_{j = 1}^k \lambda_j (\alpha \| \bm v_j \|_2^2 + (1 - \alpha) \sum_{i = 1}^n \|v_{ij}\|_1 )
    \label{eq:loss_function}
\end{align}
is minimized under the constraint that $\bm V'\bm V = \bm I_k$, which corresponds to the Stiefel Manifold, defined as the set of orthonormal matrices, i.e.,~$\text{St}(p, k) = \left \{ \bm V \in \mathbb{R}^{p\times k}: \bm V'\bm V = \bm I_k\right \}$. 

We can therefore cast this problem in the framework of optimization on manifolds. Gradient methods such as the Newton or Conjugate Gradient algorithm on manifolds have already been studied in~\citet{Edelman1998}, and \citet{Bonnabel2013} have extended SGD (Stochastic Gradient Descent) to the case when the objective function is defined on a Riemannian manifold and derived convergence properties for the algorithm. 
When the gradient is calculated on the whole dataset, it is referred to as batch gradient descent in the machine learning literature, see, e.g., \citet{Goodfellow2016}. When big amounts of data have to be processed, (minibatch) SGD is preferable, this means that the gradient is computed on each sample, or a minibatch of samples. Mathematically, the true gradient of the objective function corresponds to the expectation of the gradient over the data-generating distribution. For batch gradient descent, the true gradient is approximated by the gradient over the whole training set. When there is a large number of observations, however, computing this expectation over the whole dataset is not feasible, therefore it is computed over random subsamples, which are called minibatches \citep{Goodfellow2016}. Therefore, by application of a suitable variant of SGD we can ensure the scalability of the proposed algorithm to datasets containing a very large number of observations, which would not be possible with an approach based on alternating regressions, for example. In the following, we describe the algorithm for batch gradient descent for ease of notation.

\subsection{Manifold optimization}

Given a starting point $\bm V_0 \in \text{St}(p, k)$, subsequent iterations lie on the same manifold. This is accomplished as follows: 
First, the gradient at step $t$, $\bm G_t := \nabla_{\bm V_t} \mathcal{L}(\bm V_t) = \left ( \frac{\partial \mathcal{L}(\bm V_t)}{\partial \bm v_l}\right )_{l = 1}^k$, is computed, then the gradient is projected on the tangent space of the manifold at the current parameter value, denoted by $\mathcal{T}_{\bm V_t}\text{St}(p,k)$. Let $\bm P_{\bm G_t}$ denote the projection of the gradient, which can be computed as $\bm P_{\bm G_t} = -\gamma_t (\bm I_p - \bm V_t \bm V_t') \bm G_t$, where $\gamma_t$ refers to the step size. 

Finally, the gradient step is executed on the manifold. This can be done via an exponential map, $\bm V_{t+1} \gets \exp_{\bm V_t}(-\gamma_t \bm P_{\bm G_t})$, or via a retraction, a first-order approximation of the exponential map, denoted by $\bm V_{t+1} \gets R_{\bm V_t}(-\gamma_t \bm P_{\bm G_t})$. Numerically, the latter is preferable, and we can use $R_{\bm V_t}(\bm P) = \text{qf}(\bm P)$, where qf() extracts the orthogonal factor from the QR decomposition. This particular retraction was also studied in Proposition 3 in \citet{Bonnabel2013} and essentially follows the gradient in the Euclidean space and then orthonormalizes the matrix at each step. 

For the computation of the gradient, the continuous differentiability of the loss function $\rho$ is a necessary condition. This assumption is fulfilled for the least squares loss without regularization, for the robust loss functions and different penalties, however, it does not hold. While the Huber loss \eqref{eq:Huber} can be approximated by the differentiable Pseudo-Huber loss function \citep{Hartley2003}, $\rho_{PH}(r) = b^2(\sqrt{1 + (\nicefrac{r}{b})^2} - 1)$,  the treatment of the LTS loss \eqref{eq:LTS} requires more thought. It is easy to see though, that a gradient step in terms of the $\rho_{LTS}$ function can be expressed as a re-weighting step: $\rho_{LTS}$ is continuously differentiable on the set of points for which $|r| \leq |r|_{(h)}$. Let $H_t$ denote this $h$-subset computed based on the current iterate $\bm V_t$, then $\mathcal{L}(\bm V_t, H_t)$ corresponds to the value of the objective function depending on $H_t$. After each gradient step, $H_{t+1}$ is updated using the current iterate $\bm V_{t+1}$ and we get the inequality
\begin{align}
    \mathcal{L}(\bm V_{t+1}, H_{t+1}) \leq \mathcal{L}(\bm V_{t+1}, H_{t}) \leq \mathcal{L}(\bm V_t, H_t),
\end{align}
where the inequality on the right is due to the gradient update \citep{Bonnabel2013}, and the inequality on the left holds because the largest standardized residuals are trimmed.
Note that for the stochastic variant of the gradient step, the inequality only holds in expectation.

\subsection{Initialization}
\label{sec:initialization}

 As the loss function \eqref{eq:loss_function} is based on the approximation of the data matrix via rank-$k$ SVD, it seems natural to consider the first $k$ singular values as starting points. Despite the desirable properties of Riemannian SGD studied by \citet{Bonnabel2013}, however, we cannot hope to get convergence to a global minimum, as the loss function is not convex on $\text{St}(p, k)$. Therefore, it is crucial to choose an initial estimate $\bm V_0$ that is robust in the presence of outliers.

We propose first applying a transformation $g$ to the data matrix and then computing the SVD of $\bm Y=g(\bm X)$, resulting in the first $k$ right-singular vectors of $g(\bm X)$ as the initial estimate $\bm V_0$. 
The procedure is inspired by the robust high-dimensional product-moment correlation, studied in \citet{Raymaekers2021}
where robustness properties of different data transformations are investigated. In the following, we consider one of the following options
for the transformation, and later on, compare their results.
Both options involve estimators of location $t_j$ and scale $c_j$ of the variables ($j=1,\ldots ,p)$. Here we use the median 
for location and the Qn estimator \citep{RouC1993} for scale.
\begin{enumerate}
    \item 
    Rank transformation: The elements of the transformed data matrix
    $\bm Y$ are given by $y_{ij} = g(x_{ij}) = \nicefrac{1}{n}(\text{rank}_i( x_{ij}) - 0.5) \cdot c_j + t_j$.
    \item Wrapping transformation: Denote $z_{ij}=\frac{x_{ij}-t_j}{c_j}$, for $i = 1, \ldots, n$ and $j = 1, \ldots, p$, as the column-wise robustly standardized data.
    The elements of the transformed data matrix $\bm Y$ are given by $y_{ij} = g(x_{ij})= \psi_{b,c} (z_{ij})\cdot c_j+t_j$. The $\psi$-function is given by 
    \begin{align}
        \psi_{b, c}(z) = \begin{cases} 
         z &\text{if~} 0 \leq |z|\leq b,\\
         q_1 \tanh{(q_2(c - |z|))} \text{sign}(z) &\text{if~} b \leq |z|\leq c,\\
         0 &\text{otherwise},
       \end{cases}
    \end{align}
    where the values $q_1$ and $q_2$ can be derived for any combination of $0 < b < c$ \citep{Raymaekers2021}. We use the default values 
    $b=1.5$ and $c=4$ as proposed in~\citet{Raymaekers2021} and implemented in the function \texttt{wrap()} of the \proglang{R} package \pkg{cellWise} \citep{R_cellwise}.
\end{enumerate}
The choice of transformation is discussed in \citet{Raymaekers2021}, and depends on the dataset at hand. The application of a rank transformation is straightforward, computationally attractive for large datasets and does not require additional parameter choices. Wrapping can be tuned for a better compromise between robustness and efficiency, requiring additional computational resources.

\subsection{Residual scale}

Based on an initial estimate $\bm V_0$, we can compute the residuals
$\bm R(\bm V_0)=\bm X-\bm X \bm V_0 \bm V_0'$ with the elements 
$r_{ij}(\bm V_0)$. The objective function~\eqref{eq:loss_function}
needs an estimate of the residual scale, $\hat{\sigma}_j$, for the 
$j$-th column of this matrix ($j=1,\ldots ,p).$ Minimizing the objective function then yields an updated estimate of $\bm V$, and thus also new residuals, from which the residual scale needs to be re-estimated.
For estimating this residual scale, \citet{Maronna2008} suggest using an M estimator of scale. As the proposed algorithm requires repeated computation of the residual scale, it is desirable to choose an estimator with a high breakdown point which is also easy to compute. We therefore propose to use the simple least median of squares estimator given by $\hat{\sigma}_j = \text{median}_i|r_{ij}|$ for this purpose. 

\subsection{Sparsity inducing penalties}

When the elastic net parameter in \eqref{eq:loss_function} is set to $\alpha = 0$, we get an $L_1$-penalty, a popular choice for inducing sparsity in the PCA loadings \citep{Croux2013, Hubert2016}.
While the $L_1$-norm is not continuously differentiable, it can be approximated by a differentiable function, which converges towards the $L_1$-norm. In our implementation we use $|v_{jl}|= v_{jl} \text{sign}(v_{jl}) = \text{lim}_{c\rightarrow \infty} v_{jl}\text{tanh}(c \cdot v_{jl})$, as discussed by \citet{Oellerer2015}. The more the constant $c$ is increased, the better the absolute value function is approximated. In our implementation, we set $c = 1000$. Due to this approximation and the nature of gradient-based updates, this procedure does not lead to true sparsity in the loadings but only shrinks the elements of the loadings matrix close to zero. In order to get truly sparse results, we propose to threshold the loadings in the following way: First, we track the relative change in each of the elements of the loadings matrix between the iterations $\bm V_t$ and $\bm V_{t+1}$. Then, the threshold value is computed as the average change of all elements during the last $M$ iterations plus two standard deviations. If the absolute value of an entry of $\hat{\bm V}$ is lower than the threshold, it is set to 0. In the implementation, we use $M = 10$, which we found to be a reasonable choice in our experiments, as increasing $M$ above this value brought minimal improvement.

The complete algorithm is summarized in Algorithm \ref{alg:sgd_manifold}.

\begin{algorithm}
	\caption{Robust and Sparse PCA via Manifold Optimization}\label{alg:sgd_manifold}
	\begin{algorithmic}[1]
        \State Compute transformations $\bm Y \gets g(\bm X)$
        \State Initialize $\bm V_0$ as first $k$ right-singular vectors of $\bm Y$
        \While {$\|\mathcal{L}(\bm V_{t+1}) - \mathcal{L}(\bm V_{t})\| > \delta$}
    	\State $\bm G_t \gets \nabla_{\bm V} \mathcal{L}(\bm V_t)$ \Comment{compute gradient H}
            \State $\bm P_{\bm G_t} \gets \text{proj}_{\mathcal{T}_{\bm V_t}\text{St}(p,k)}(\bm G_t)$ \Comment{projection of gradient onto tangent space}
            \State $\bm V_{t+1} \gets R_{\bm V_t}(-\gamma_t \bm P_{\bm G_t})$ \Comment{gradient step via retraction}
            \If {$\rho = \rho_{LTS}$}
                \State update $H_{t+1}$ based on $\bm V_{t+1}$  \Comment{update $h$-subset}
            \EndIf
            \State $ d_{t+1} \gets \nicefrac{\|\bm V_{t+1} - \bm V_t\|}{\|\bm V_t\|}$ \Comment{track relative change}
       \EndWhile
       \State $\bar{t} \gets \text{avg}[d_m]_{m=i}^{i-M+1} + 2\text{sd}[d_m]_{m=i}^{i-M+1}$ \Comment{compute threshold}
       \State $ \hat{\bm V}\gets [ v_{jl} \text{~if~} | v_{jl}| > \bar{t}, 0 \text{~otherwise}]_{jl}$ \Comment{thresholding}\label{alg:thresh_a}
       \State $\hat{\bm Z} \gets \bm X \hat{\bm V}$ \Comment{compute robust scores}
       \State $\hat{\bm a} \gets \text{Qn}^2(\bm X \hat{\bm V})$ \Comment{compute robust variances of the scores}
	\end{algorithmic} 
\end{algorithm}

Step 14 of Algorithm~\ref{alg:sgd_manifold} returns the principal components, and Step 15 their variances, here estimated per component with the Qn-scale estimator~\citep{RouC1993}.

\subsection{Selection of sparsity parameter}
While the elastic net mixing parameter $\alpha$ in \eqref{eq:loss_function} is set in advance by the user, the sparsity parameter $\lambda$ is chosen depending on the data. Similarly to \citet{Croux2013}, we choose the tradeoff-product criterion (TPO) that leads to a compromise between explained variance and sparsity in the loadings. In the following, we denote the columns of the estimate $\hat{\bm V}$ as $\hat{\bm v}_l$, for $l = 1, \ldots, k$.
The original criterion 
\begin{align*}
\text{TPO} = \sum_{l = 1}^k\text{Qn}^2(\bm X \hat{\bm v}_l) \cdot \left(1 - \frac{\#\{\hat{\bm v}_l\neq 0\}}{p}\right)
\end{align*} 
where $\#\{\hat{\bm v}_l\neq 0\}$ returns the number of non-zero components in $\hat{\bm v}_l$,
can be adapted for non-sparse regularization by including the elastic net parameters $\alpha$:  
\begin{align}
\text{TPO} = \sum_{l = 1}^k\text{Qn}^2(\bm X \hat{\bm v}_l) \cdot \left(1 - \alpha \frac{\#\{\hat{\bm v}_l\neq 0\}}{p}\right). \label{eq:TPO}
\end{align} 

The maximum of this score function is now determined using Bayesian optimization: The advantage of this procedure is that, contrary to cross-validation in combination with grid search, the information from previous function evaluations can be exploited. This way, a bigger search space can be covered. For the basic algorithm, it is assumed that there is a budget of in total $N$ function evaluations. A Gaussian prior is placed on the score function, then its value is observed at $n_0$ points. Until $N$ iterations are reached, the following steps are repeated: (i) update the posterior probability distribution on the score function, (ii) determine the maximum of the acquisition function, and (iii) observe the score value at this parameter configuration.
 We used the implementation in the \proglang{R} package \pkg{ParBayesianOptimization} \citep{R_Wilson2022} with the commonly used expected improvement as the acquisition function and the tradeoff product optimization (TPO) as the score function.

The proposed algorithm is called SCRAMBLE (Sparse Cellwise Robust Algorithm for Manifold-based Learning and Estimation) and implemented in \proglang{R} \citep{R_Rlanguage}, in package 
\pkg{scramble}, available in the supplement.

\section{Simulations}
\label{sec:sim}

A simulation study was conducted to evaluate the performance of the proposed method using different loss functions in comparison with other approaches, namely Robust and Sparse PCA (ROSPCA) \citep{Hubert2016} for the rowwise contamination model, and MacroPCA \cite{Hubert2019} for the cellwise contamination model, although this method cannot induce sparsity. 

\subsection{Simulation settings}
Similar to \citet{Croux2013, Hubert2016, Hubert2019}, several different contamination settings are considered. 
The casewise Tukey-Huber contamination model can be formalized as
\begin{align}
    X = (1-B)Y + BZ,
\end{align}
where $X$, $Y$ and $Z$ are $p$-dimensional vectors, $Y\sim F$ with $F$ corresponding to the model distribution and $Z\sim G$ with $G$ corresponding to the outlier-generating distribution. $B\sim Bin(1, \varepsilon)$ can be interpreted as a \emph{contamination indicator} \citep{Alqallaf2009}. 
In the multivariate setting, this model has been criticized for only allowing rowwise contamination, when in reality -- especially in high-dimensional settings -- it is likely that only a few columns in many rows are affected by outliers.

For this cellwise contamination framework, \citet{Alqallaf2009} have formalized the independent contamination model as
\begin{align}
    X = (\bm I - \bm B)Y + \bm B Z,
\end{align}
where $\bm B = diag(B_1, B_2, \ldots, B_p)$ are independent $B_i\sim Bin(1, \varepsilon_i)$.

A low- and a high-dimensional simulation setting is considered, and they are adapted from \citet{Croux2013} and \citet{Hubert2016}. Clean data $\bm X$
are generated from a multivariate normal distribution $\mathcal{N}_{p}(\bm 0, \bm \Sigma)$ with zero means and covariance $\bm \Sigma = \bm C^{1/2} \bm A \bm C^{1/2}$ given as follows :
\begin{enumerate}
    \item Low-dimensional, order 2: $p = 10, n = 50$ observations \label{sim:setting1}
    \begin{align*}
             \bm A &= \left[ {\begin{array}{ccc}
             \bm S^1_{4\times 4} &   \bm 0_{4\times 4} & \bm 0_{4\times 2} \\
              \bm 0_{4\times 4} & \bm S^2_{4\times 4}  & \bm 0_{4\times 2}\\
              \bm 0_{2\times 4} & \bm 0_{2\times 4} & \bm I_{2\times 2} \end{array} } \right],
    \end{align*}
    where \begin{align*}
    \bm S_{ij}^1 = \begin{cases} 
         1 &\text{if~} i = j,\\
         0.9 &\text{otherwise},
       \end{cases}
       \end{align*} and
    \begin{align*}
    \bm S_{ij}^2 = \begin{cases} 
         1 &\text{if~} i = j,\\
         0.7 &\text{otherwise}.
       \end{cases}
       \end{align*}
    The matrix $\bm C$ assigns the same scale to all variables of the same group, and it is given as $\bm C = \text{diag}(\bm {100}_4, \bm {25}_4, \bm 4_2)$, where $\bm a_b$ is a vector with $b$ replicates of the number $a$. 
    The true loadings of the first component in this setting are $\bm v_1 = \nicefrac{1}{2}(\bm 1_4, 0, \ldots, 0)$, and of the second component $\bm v_2 = \nicefrac{1}{2}(\bm 0_4, \bm 1_4, 0, \ldots, 0)$. 
    
    \item High-dimensional, order 2: $p = 500, n = 100$ observations
    \begin{align*}
             \bm A &= \left[ {\begin{array}{ccc}
             \bm S^1_{20\times 20} &   \bm 0_{20\times 20} & \bm 0_{20\times 460} \\
              \bm 0_{20\times 20} & \bm S^2_{20\times 20}  & \bm 0_{20\times 460}\\
              \bm 0_{460\times 20} & \bm 0_{460\times 20} & \bm I_{460\times 460} \end{array} } \right],
    \end{align*}
    where $\bm S^1$ and $\bm S^2$ are defined as before. $\bm C$ is given as $\bm C = \text{diag}(\bm {100}_{20}, \bm {25}_{20}, \bm 4_{460})$. Similarly, the true loadings for the first 2 components correspond to the vectors $\bm v_1 = \nicefrac{1}{\sqrt{20}}(\bm 1_{20}, 0, \ldots, 0)$ and $\bm v_2 = \nicefrac{1}{\sqrt{20}}(\bm 0_{20}, \bm 1_{20}, 0, \ldots, 0)$.
\end{enumerate}

Casewise contaminated data are generated by replacing $\varepsilon \%$ of rows with data generated from a $p$-variate normal distribution $\mathcal{N}(\bm \mu_{out}, \bm I_p)$, with $\bm \mu_{out} = (2,4,2,4,0,-1,1,0,1,-1, \ldots, 1,0,1,-1)$, corresponding to the $\bm \mu_{out}$ vector suggested by \citet{Croux2013} and also used by \citet{Hubert2016}. 
For the cellwise contamination setting, data are generated using the \texttt{generateData(outlierType = "cellwiseStructured")} function in the \proglang{R} package \pkg{cellWise} \citep{R_cellwise}, where $\varepsilon \%$ of cells are replaced by multiples of the last eigenvector of $\bm \Sigma$, restricted to the dimensions of the contaminated cells \citep{Agostinelli2015, Rousseeuw2018}.

\subsection{Performance measures}

To measure the accuracy of the algorithm, we use the \emph{principal angles} between subspaces, which are defined as follows: Let $\bm V$ and $\hat{\bm V}$ be orthonormal bases for the subspaces $\mathcal{V}$ and $\hat{\mathcal{V}}$, and assume that $\text{dim}(\mathcal{V}) \leq \text{dim}(\hat{\mathcal{V}})$. Then, the principal angle $\theta$ can be computed as $\theta = \sin^{-1}(\sigma_{\max}((\bm I - \bm V \bm V')\hat{\bm V}))$, where $\sigma_{\max}$ corresponds to the largest singular value of the projected matrix \citep{Bjorck1973}. We report the principal angle scaled to lie in [$0,1$], as implemented in the function \texttt{angle()} in the \proglang{R} package \pkg{rospca} \citep{R_rospca} and described in \citet{Hubert2005}. 

The correct level of sparsity is evaluated by the true-positive rate (TPR) and true-negative rate (TNR), defined as 
\begin{align}
     \text{TPR}(\bm V, \hat{\bm V}) &=\frac{\#\{(j,k): v_{jk} \neq 0 \text{~and~} \hat{v}_{jk} \neq 0\}}{\#\{(j,k): v_{jk} \neq 0\}} \\
     \text{TNR}(\bm V, \hat{\bm V}) & =\frac{\#\{(j,k): v_{jk} = 0 \text{~and~} \hat{v}_{jk} = 0\}}{\#\{(j,k): v_{jk} = 0\}},
\end{align}
where $v_{jk}$ and $\hat{v}_{jk}$ are the corresponding elements of
$\bm V$ and $\hat{\bm V}$, respectively.
TNR and TPR correspond to the rate of correctly identified non-zero elements and zero elements, respectively, in the loadings.

The scaled principal angle corresponds to the accuracy of the subspace estimation and should be controlled at a low level even in the presence of outliers. The TPR and TNR depict how stable the correct estimation of the sparsity is; these performance measures should be close to 1.

\subsection{Simulation results}
The proposed method is compared to ROSPCA \citep{Hubert2016} in the casewise contamination setting, using the implementation in the \proglang{R} package \pkg{rospca} \citep{R_rospca}, and MacroPCA as implemented in the \proglang{R} package \pkg{cellWise} \citep{R_cellwise} in the cellwise contamination setting. While the latter does not enforce sparsity in the loadings, it is the only other cellwise robust PCA method that software is readily available for, and a comparison in terms of accuracy of subspace estimation should be insightful.

We provide results on two simulation exercises, one which investigates the runtime of several methods and a second one which compares the performance of different algorithms in terms of subspace estimation and sparsity detection.

\subsubsection{Runtime}
In addition to the algorithms ROSPCA and MacroPCA, we also include
the PP algorithm from the \pkg{pcaPP} package \citep{R_pcaPP} as another sparse and robust PCA method for high-dimensional data in the comparison. However, this method is not included in the performance
study, as the detailed comparisons to ROSPCA in \citet{Hubert2016} resulted in worse performance than the latter. For SCRAMBLE, the wrapping transformation and the Huber loss function were used.
Here we use Setting \ref{sim:setting1} described above, keeping the sparsity parameters for all methods fixed and varying $p$ and $n$. The computation was repeated 100 times, and in Figure \ref{fig:comparison_runtime} the means are reported. In the left part of Figure \ref{fig:comparison_runtime}, the number of variables is fixed at $p = 10$ and the number $n$ of observations increased, on the right side, the number of observations is fixed at $n = 50$ and the number of variables $p$ increased.

Figure~\ref{fig:comparison_runtime} shows in most situations a
higher runtime of the proposed method (SCRAMBLE). 
This is due to the other algorithms being implemented in C++. 
Still, with growing $n$ and $p$, the advantage of the proposed method becomes apparent. If the number of observations $n$ increases, the runtime even decreases, as the initial estimate is better, leading to faster convergence. In addition, very large $n$ can also be handled by an appropriate SGD variant, as the data can be processed in batches of suitable sizes, a clear advantage in comparison to the repeated subset evaluations that need to be done for the ROSPCA algorithm, for example. 
When the number of variables $p$ grows, the approach based on gradient descent also scales better, as several directions can be computed at once and no repeated cycling of candidate directions is necessary as for PP.  While the runtime of MacroPCA as another cellwise robust PCA method is appealing, it is not able to include a regularization and produce sparsity in the loadings.

The computational complexity of the proposed algorithm depends on three steps: 1. the data transformation, 2. the SVD for the starting value, and 3. the gradient algorithm. For the rank transformation, this results in a complexity of $O(n\log(n) + np\min(n,p) + inp + pk^2)$, and for the wrapping transformation in $O(np + np\min(n,p) + inp + pk^2)$, where $i$ refers to the number of iterations and $k$ to the number of estimated components (refer to \citet{Raymaekers2021} for the complexity of the wrapping transformation and \citet{Cunningham2015} for a discussion of complexity of Riemannian gradient descent). The learning rate was set to $\gamma = 0.001$ with a learning rate decay of $0.99$ for batch gradient descent. The simulations were run on an Apple M1 MacBook Pro 2020 (16 GB).

\begin{figure}[!htp]
\centering

\subfloat{\includegraphics[width = 1\textwidth]{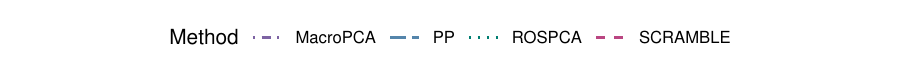}}

\subfloat{\includegraphics[width = 0.495\textwidth]{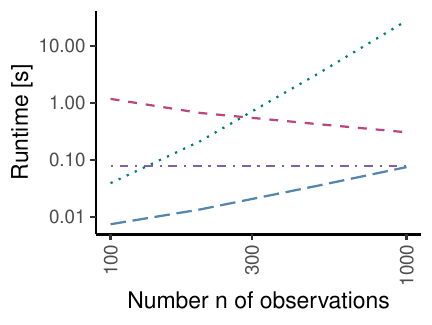}}
\subfloat{\includegraphics[width = 0.495\textwidth]{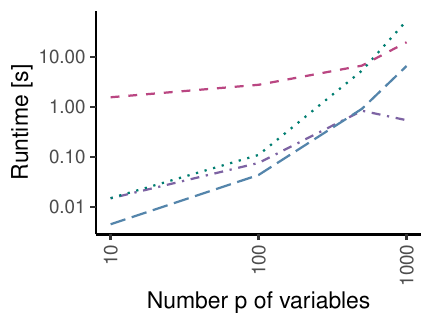}}

\caption{
Comparison of runtime for the different methods. Left: The number of variables is fixed at $p=10$, the number of observations $n$ is increased from $100$ to $1000$. Right: The number of observations is fixed at $n = 50$, and the number of variables $p$ is increased from $10$ to $1000$. For SCRAMBLE, the wrapping transformation and the Huber loss function were used.}
    \label{fig:comparison_runtime}
\end{figure}

\subsubsection{Performance}

The robustness of the methods for an increasing proportion of casewise and cellwise outliers is studied. 
For the proposed method, performance measures of different combinations of the initial transformation and loss function are evaluated and compared to a suitable alternative method. 

Figure~\ref{fig:comparison_methods_casewise} presents the results for casewise contamination. The performance results for the low-dimensional setting ($p=10$ and $n=50$) are shown in the plots on the left
side, while those for the high-dimensional setting ($p=500$ and $n=100$) are on the right-hand side. In both cases, $k=2$ is fixed, and ROSPCA is compared to different versions of SCRAMBLE, with different loss functions $\rho$ (Huber, Tukey, LTS), see Section~\ref{sec:method}, and different transformations (rank, wrapping) for the initialization, see Section~\ref{sec:initialization}.

In the low-dimensional setting, the SCRAMBLE method clearly performs better than ROSPCA in all performance measures, especially in estimating the sparsity, described by the TNR. 
The increasing contamination has only little effect on the outcome.
Also the different versions of SCRAMBLE show very similar results.
For the high-dimensional setting we can see more effects. In the uncontaminated case, SCRAMBLE outperforms ROSPCA, particularly for the TNR.
Interestingly, the TNR improves for ROSPCA with increasing contamination, and the same holds for SCRAMBLE based on the rank transformation. 
However, here an angle of about 0.75 for 20\% contamination already suggests a solution very different from the target.
The best results are achieved by SCRAMBLE initialized with the wrapping transformation, for the LTS and the Tukey loss function.

\begin{figure}[!htp]
\centering
\subfloat{\includegraphics[width = 1\textwidth]{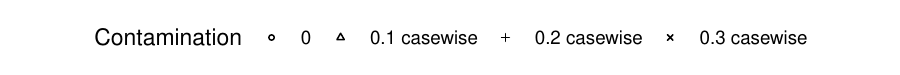}}

\subfloat{\includegraphics[width = 0.495\textwidth]{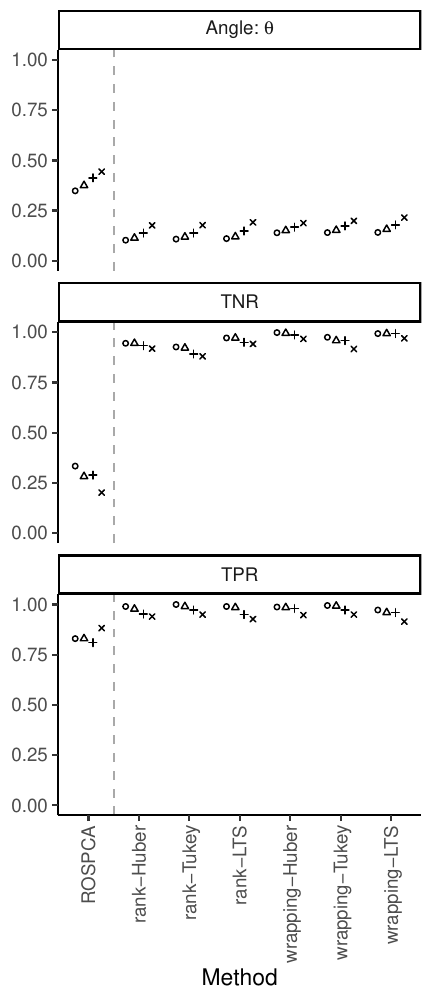}}%
\subfloat{\includegraphics[width = 0.495\textwidth]{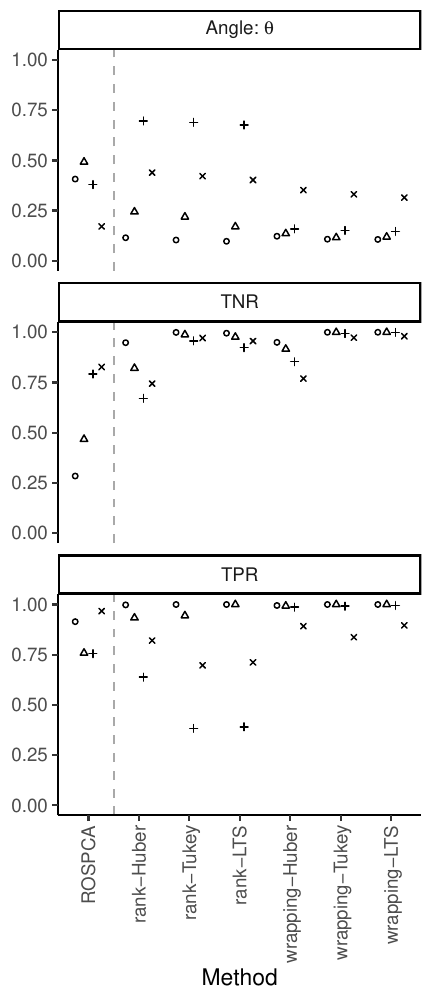}}

\caption{Comparison of performance measures for the methods 
ROSPCA and SCRAMBLE (six combinations of loss functions and transformations) for the casewise setting. Left: low-dimensional ($p=10$, $n=50$);
right: high-dimensional ($p=500$, $n=100$). The different point shapes correspond to different contamination levels: $\epsilon=0\%, 10\%, 20\%, 30\%$.}
    \label{fig:comparison_methods_casewise}
\end{figure}

The results for the cellwise contamination setting are shown in Figure \ref{fig:comparison_methods_cellwise}, with the low-dimensional setting on the left and the high-dimensional setting on the right.
Here we compare SCRAMBLE to MacroPCA as an alternative algorithm for cellwise robust PCA, although this method does not allow for sparsity. 
Consequently, MacroPCA fails to accurately predict the sparsity (resulting in a TPR of 1 and a TNR of 0) but yields very good results for the principal angle in the low-dimensional setting. 
Note that we selected the same proportions of contamination as for the
casewise setting. 
However, in the cellwise setting, these amounts contaminate many more rows of the data matrix, or even all rows~\citep{Raymaekers2023}, leading to a faster decrease in performance among all metrics.
For the high-dimensional setting, however, the proposed algorithm yields the best overall results.
This is because SCRAMBLE processes the loss function cellwise, and thus both more rows or more columns yield more training set observations.
As expected, there is a certain decrease in performance for increased contamination. 
In this setting, rank-based initialization is more robust than initialization with the wrapping transformation with default values~\citep{Raymaekers2021}. The overall best results are in
combination with the LTS or Tukey loss function.

\begin{figure}[!htp]
\centering
\subfloat{\includegraphics[width = 1\textwidth]{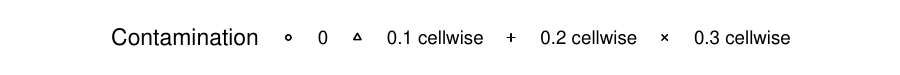}}

\subfloat{\includegraphics[width = 0.495\textwidth]{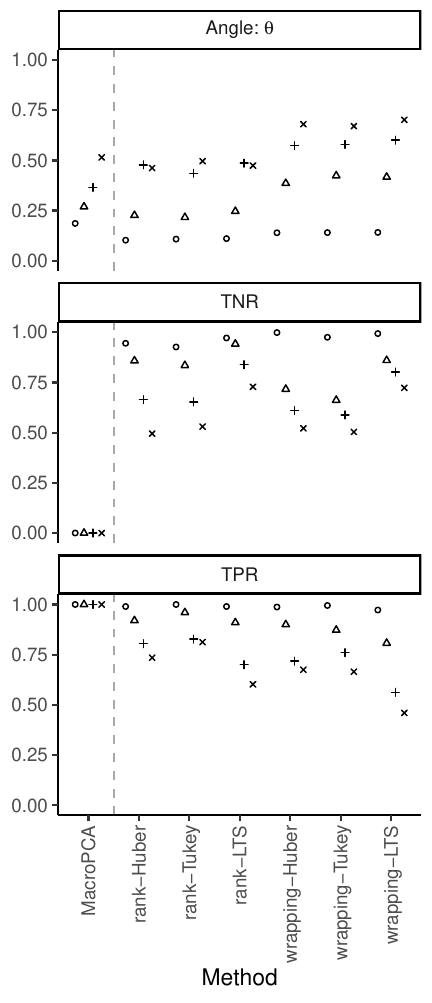}}
\subfloat{\includegraphics[width = 0.495\textwidth]{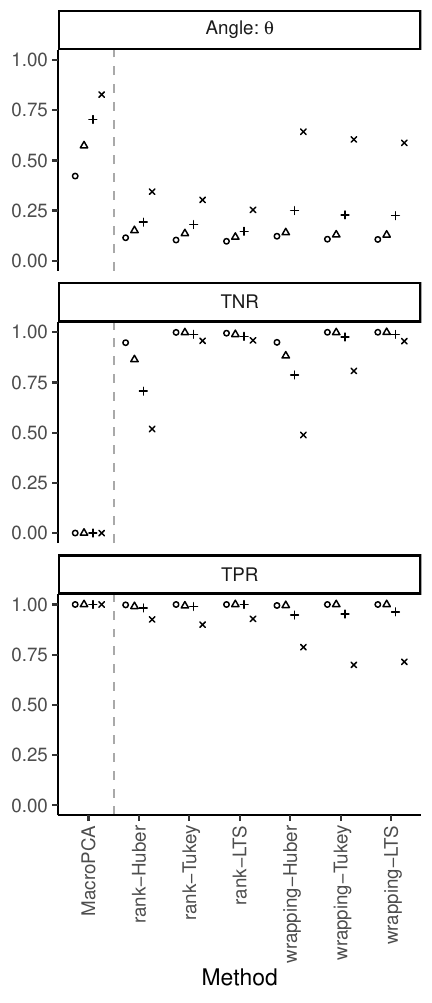}}

\caption{Comparison of performance measures for the methods 
ROSPCA and SCRAMBLE (six combinations of loss functions and transformations) for the cellwise setting. Left: low-dimensional ($p=10$, $n=50$);
right: high-dimensional ($p=500$, $n=100$). The different point shapes correspond to different contamination levels: $\epsilon=0\%, 10\%, 20\%, 30\%$.}
    \label{fig:comparison_methods_cellwise}
\end{figure}

\section{Illustration on real data}
\label{sec:exa}

The usefulness of the approach will be demonstrated on two datasets from tribology, the study of friction, wear, and lubrication. The presented data originate from chemical analyses and tribological experiments performed on automotive engine oils after they have been subjected to a varying duration of artificial alteration in the laboratory, see \citet{Doerr2019} and \citet{Besser2019} for a description of the alteration methods. FTIR (Fourier-transform infrared) spectra consist of absorption values that are measured over about $2000$ wavenumbers, with distinctive peaks associated with certain oil attributes. For this data structure, the sparsity assumption can be justified: It can be assumed that only a small set of wavenumbers is sufficient to explain most of the variability in the dataset \citep{Pfeiffer2022}. 
Another aspect is lubrication performance, which can be measured on an Schwing-Reib-Verschleiss (SRV) tribometer experiment (a steel ball sliding against a steel disk with the lubricant of interest in between, see \citet{Doerr2019} for a more detailed description of the experiment). One part of the resulting data consists of optical images (taken under the microscope) of the wear scar areas, which yield data matrices with $n \ll p$ when vectorized.

As both types of data are produced in the laboratory, we can expect outliers to be present in the datasets due to possibly high variability following the alteration process and experimental effects. In addition, the experiments are often costly and time-consuming, resulting in much fewer observations than variables and wide data matrices. Thus, dimension reduction is necessary before applying any further analysis. We demonstrate in the following how the proposed method can be applied to perform this dimension reduction via robust PCA and, in addition, yield sparse loadings when appropriate.

\subsection{FTIR spectra}
The presented dataset consists of $n= 50$ FTIR spectra of 10 automotive engine oils. The fresh oils were subjected to a small-scale alteration in the laboratory as described in \citet{Doerr2019}. During the alteration, samples were taken regularly and spectra were recorded, each containing the absorbance at $p = 1668$ wavenumbers. 
The task at hand is to understand which variables contribute most to the variability in the dataset, and often classical PCA is applied, see, for example, \citet{Besser2013}. To make it more challenging, we contaminate the dataset with $6$ observations originating from a large-scale alteration \citep{Besser2019} to imitate a scenario when the origin of a sample may not be as clear as in the laboratory setting.
Our aim is to identify observations that are different from the majority of the data,
and also understand why they are outlying. As mentioned before, sparsity in the PCA loadings is desirable in this setting to enhance interpretability. 

The resulting dataset consists of $n = 56$ observations and $p = 1668$ variables. As spectral data are already on the same scale, the data are not scaled, but only column-wise centered with the median before applying the methods. We compare the results from ROSPCA \citep{Hubert2016} and the proposed SCRAMBLE method with the rank-based data transformation for the starting value and the Huber loss function. Due to the high number of variables in the dataset, the rank-based transformation is advisable. The different loss functions performed similarly, therefore we only show the results for the Huber loss. For both methods, the number of principal components is determined via the cumulative proportion of explained variance. For ROSPCA, this results in $k_{\text{ROSPCA}}=10$, for SCRAMBLE in $k_{\text{SCRAMBLE}} = 7$ components. Then, hyperparameter optimization for the sparsity parameters is done for both algorithms. For SCRAMBLE, this is done using Bayesian optimization and the TPO criterion; for RSPCA, the grid search and BIC type criterion proposed by \citet{Hubert2016} is used.
Figure~\ref{fig:loadings_ftir} shows the original FTIR spectra in gray, together with the first (left plot) and second (right plot) loadings vectors of both methods. We find that SCRAMBLE leads to more sparsity, and therefore to results that are easier to interpret. Some of the selected variables are known to be associated with underlying chemical processes during oil alteration, like oxidation, conventionally evaluated at wavenumber $1720$ cm$^{-1}$, or phenolic antioxidants at $3650$ cm$^{-1}$ \citep{Besser2019}.

\begin{figure}[htp]
\centering
\subfloat{\includegraphics[width = 1\textwidth]{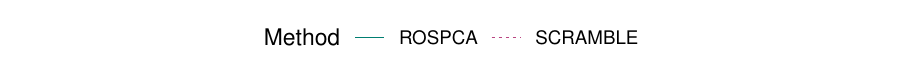}}

\subfloat{\includegraphics[width = 0.495\textwidth]{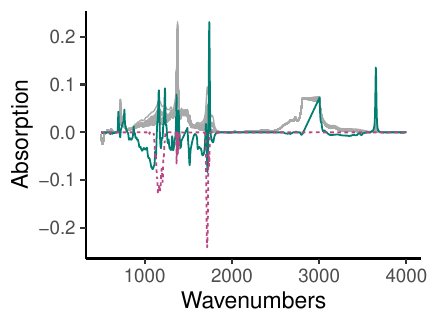}}
\subfloat{\includegraphics[width = 0.495\textwidth]{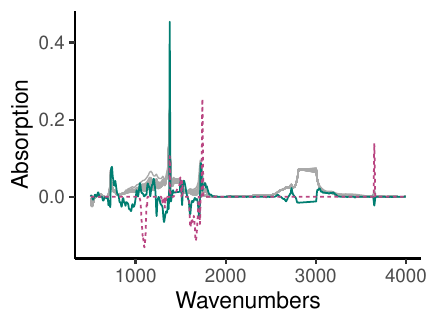}}
\caption{FTIR spectra of the original data, shown in gray,
together with the first (left plot) and second (right plot) loadings vectors of each method.
}
    \label{fig:loadings_ftir}
\end{figure}

Figure \ref{fig:distance_distance_ftir} shows PCA diagnostic plots based on the score distance SD and orthogonal distance OD, with the outlier cutoff values as dashed lines, see~\citet{Hubert2005}.
A high value of the OD means that the observations are far away from the 
estimated PCA subspace. The left plot presents the results for ROSPCA, and here almost all outliers, i.e.,~the observations from large-scale alteration, are identified with high OD values. However, also four regular observations yield high OD values. The right plot for the SCRAMBLE results corresponds to what we would expect.

\begin{figure}[htp]
\centering
\subfloat{\includegraphics[width = 1\textwidth]{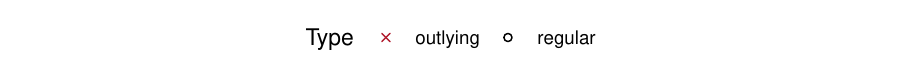}}

\subfloat{\includegraphics[width = 0.495\textwidth]{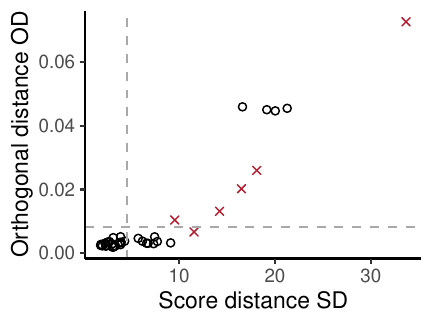}}
\subfloat{\includegraphics[width = 0.495\textwidth]{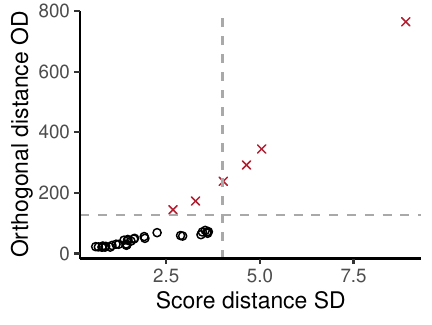}}
\caption{Score distance versus orthogonal distance for the FTIR spectra. Results for ROSPCA (left) and SCRAMBLE (right).}
    \label{fig:distance_distance_ftir}
\end{figure}

In order to investigate the differences between ROSPCA and SCRAMBLE in more detail, we show plots of the standardized residuals in Figure~\ref{fig:residuals_ftir} for a selected wavenumber range and a subset of the observations.
The left plot is for ROSPCA, the right plot is for SCRAMBLE, and each tile represents one element in the scaled residual matrix, with color according to the legend. The residuals were standardized robustly using the median and mad of the residual matrix. Note that the residuals scale is very small leading to very large scaled residuals for some cells.
The first six rows correspond to the FTIR spectra of oils from a different alteration process, and SCRAMBLE clearly reconstructs these worst, meaning they are not as influential to the fit of the PCA subspace. In ROSPCA, on the other hand, only observation $6$ is the only clearly visible observation out of the outlier subset, and four further observations also show larger residuals. A look at the original spectra with a zoomed-in view into this wavenumber range in Figure~\ref{fig:zoom_abs} explains this behavior: There, the outliers (first 6 rows) are shown by red dashed curves, and only one (observation 6) is clearly further away, while the other outliers partially overlap with regular observations. This overlap around wavenumber 1740 cm$^{-1}$, thus in a very restricted range, is the reason why four regular observations are falsely classified as outlying by the ROSPCA algorithm. The affected wavenumbers lie in the absorption band of oxidation products, ranging from 1860--1660 cm$^{-1}$, and are of interest in conventional analysis of FTIR spectra \citep{Besser2019, Pfeiffer2022}. A cellwise robust method can assist the practitioner in finding and understanding the differences between outlying and regular observations.

\begin{figure}
\centering
\begin{adjustbox}{minipage=\textwidth, scale = 1}
\subfloat{\includegraphics[width = 1\textwidth]{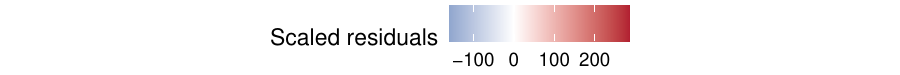}}

\subfloat{\includegraphics[width = 0.495\textwidth]{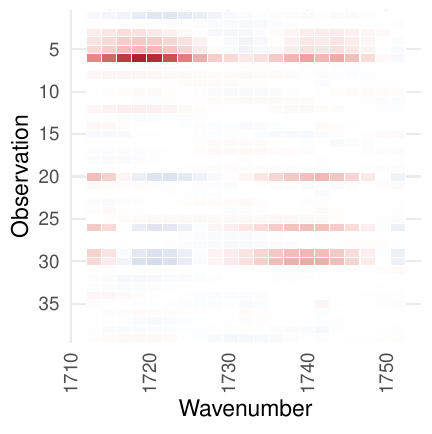}}
\subfloat{\includegraphics[width = 0.495\textwidth]{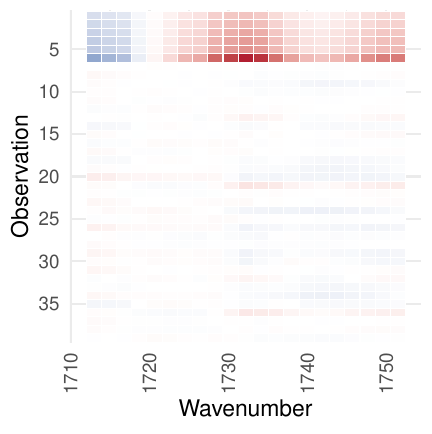}}
\caption{Residual plots for a selected range of wavenumbers and for a subset of the observations. Left: ROSPCA residuals; right: SCRAMBLE residuals. }
    \label{fig:residuals_ftir}
    \end{adjustbox}
\end{figure}

\begin{figure}[htp]
\centering
\includegraphics[width = 0.495\textwidth]{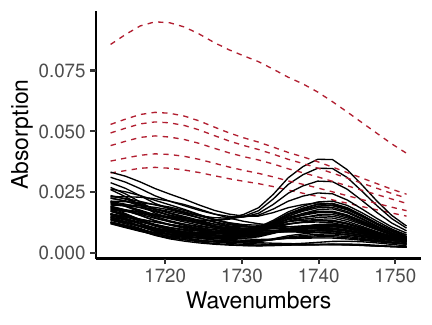}
\caption{Zoomed-in view of the FTIR spectra from Figure~\ref{fig:loadings_ftir} for the selected wavelength range shown in Figure~\ref{fig:residuals_ftir}. The red dashed lines are outliers with large-scale alteration.}
\label{fig:zoom_abs}
\end{figure}

In summary, we can see the benefit of a cellwise robust method in contrast to only casewise robust estimation. As the differences only show at certain peaks in the spectra, a cellwise robust method does not lose as much information as a casewise method, resulting in a better fit with fewer components. In addition, we can also identify the variables which contribute most to the outlyingness.

\subsection{Tribology: wear scar images}

Often, the task is not only to derive useful features but also to predict an outcome or a property. In this example, we demonstrate the flexibility of our approach in a PCR (Principal Component Regression) setting with a wide data matrix, which calls for dimension reduction.
In \citet{Pfeiffer2023}, image features were derived from a similar image dataset, before robust regression methods were applied. For this demonstration, we derive robust features via SCRAMBLE directly from the vectorized images before applying a robust regression on the resulting principal components. We do not use a sparsity-inducing regularization, as this has been found to not provide an advantage for image data \citep{Pfeiffer2023}. 
In the given setting, $n = 220$ gray-scale images of size $64\times 64$, resulting in vectors of size $64^2 = 4096$, together with a response variable containing the alteration duration (in hours) of the lubricant used in the SRV experiment, are available. After the removal of all constant columns, $p = 3025$ columns are left. As the response variable, the alteration duration is given in hours, which is square-root transformed before estimating the model. 

We compare classical PCA via SVD to the SCRAMBLE algorithm with rank-based preprocessing and the Huber loss. Therefore, the dataset is randomly split into a 70\% training, a 20\% validation, and a 10\% test set. The principal components are estimated on the training set, and then the optimal number of components is evaluated for the validation set via the mean squared error of prediction (MSEP) using least-squares regression for the classic estimation and robust regression (the function \code{lmrob()} from the \pkg{robustbase} \proglang{R} package \citep{R_robustbase}) for robust estimation. Finally, the MSEP is computed for the test set. The first three estimated loadings are shown in 
Figure~\ref{fig:loadings_wear_classioal} for classical PCA and 
in Figure~\ref{fig:loadings_wear_robust} for robust PCA using the SCRAMBLE algorithm. We can observe that the first loadings look quite similar, while the order is different. Both methods distinguish between the border and the interior of the wear scar, as well as the overall contributions.
\begin{figure}
\centering
\subfloat{\includegraphics[width = 0.33\textwidth]{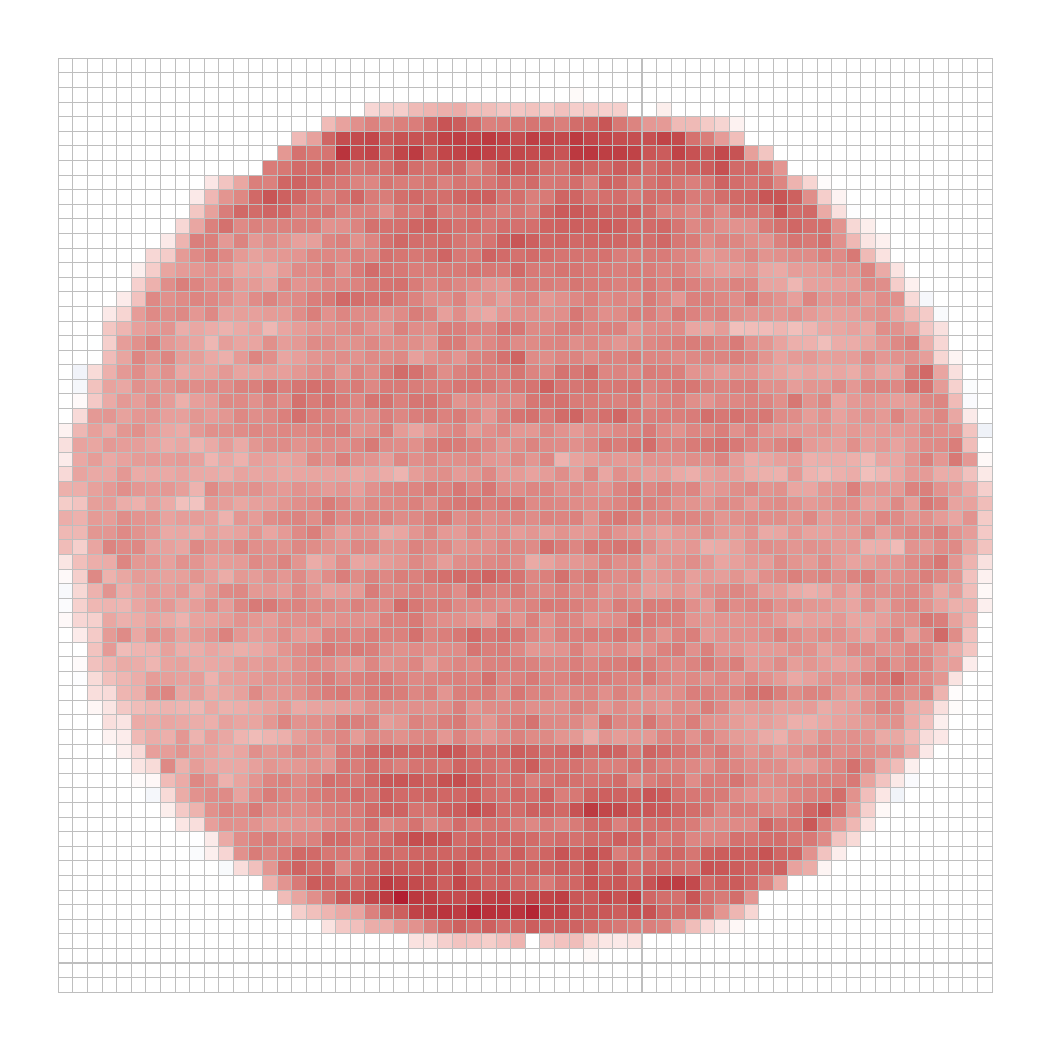}}
\subfloat{\includegraphics[width = 0.33\textwidth]{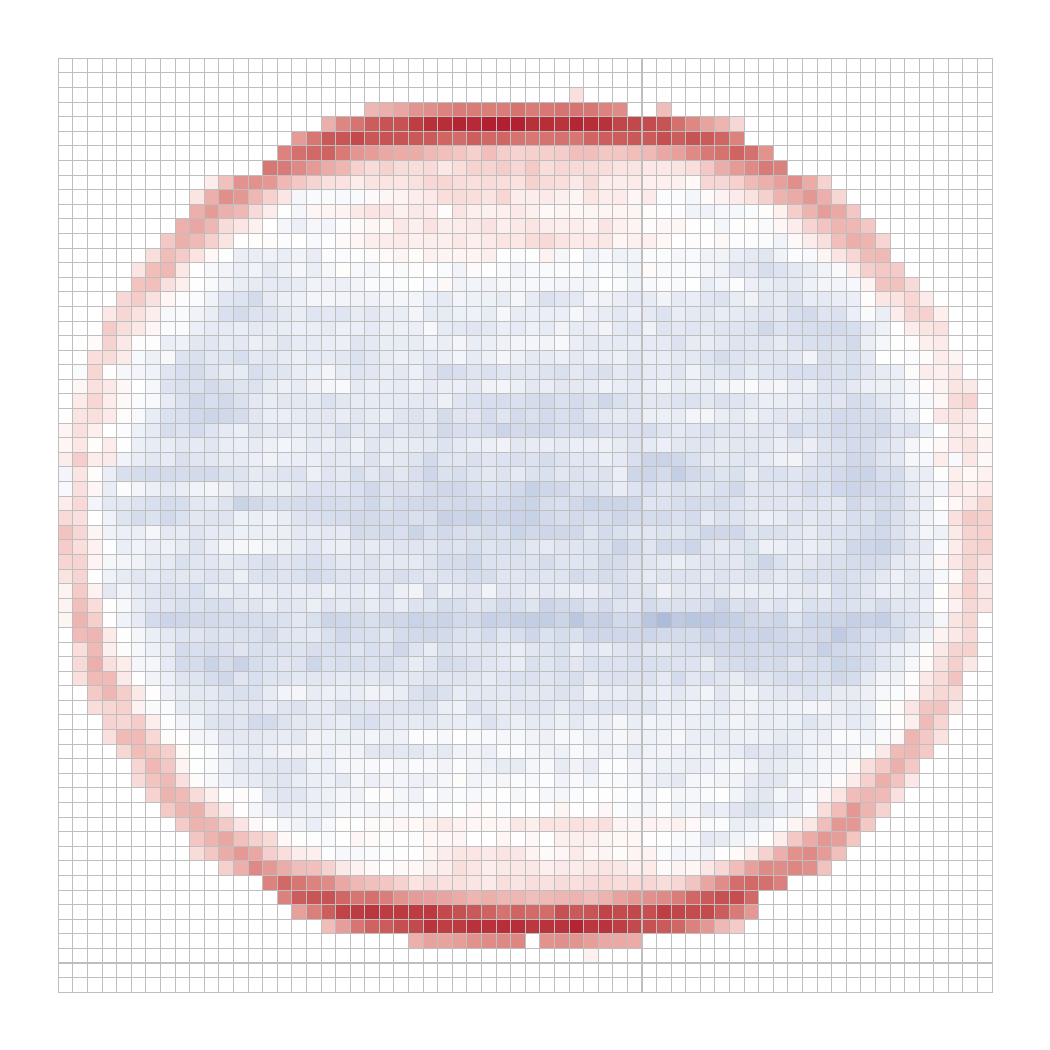}}
\subfloat{\includegraphics[width = 0.33\textwidth]{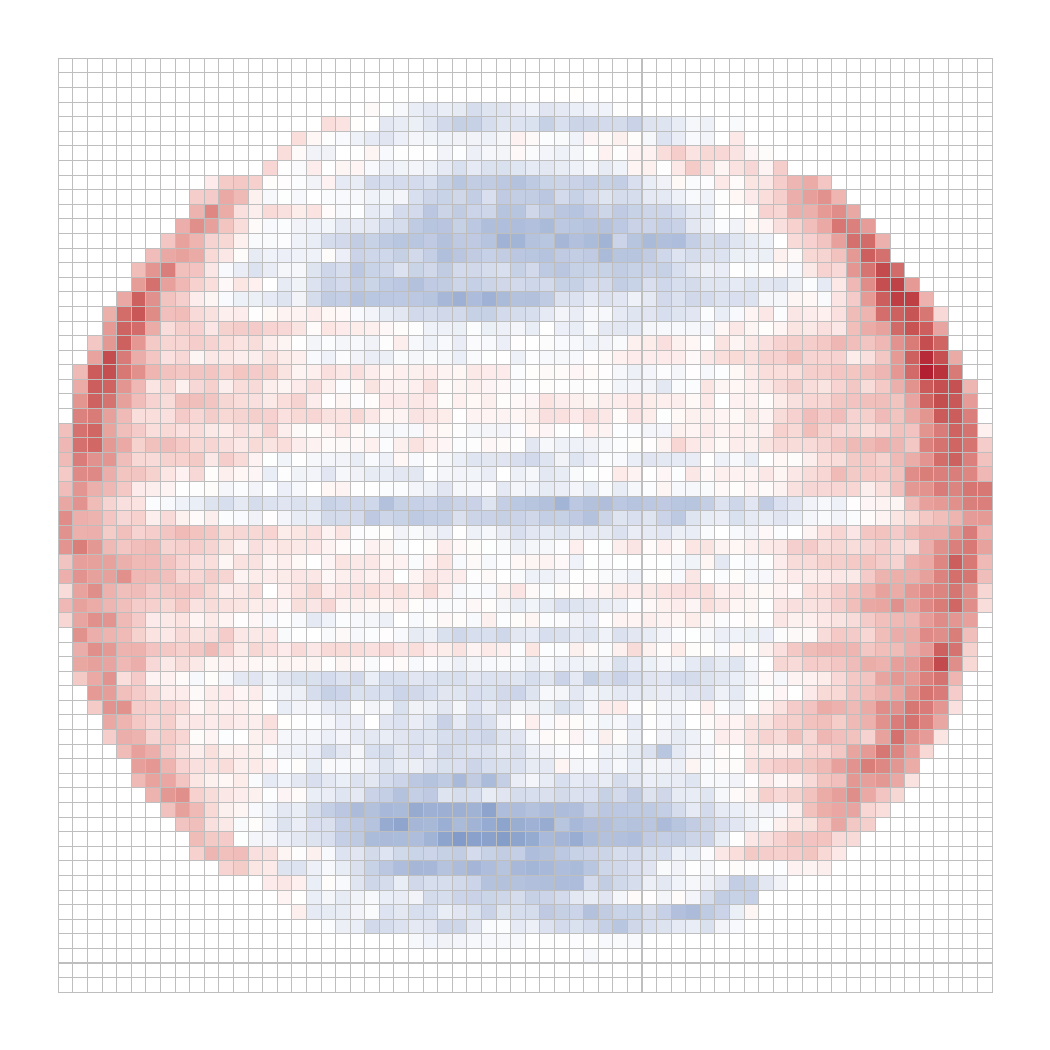}}
\caption{Classical loadings back-transformed to the image space.}
    \label{fig:loadings_wear_classioal}
\end{figure}
\begin{figure}
\centering
\subfloat{\includegraphics[width = 0.33\textwidth]{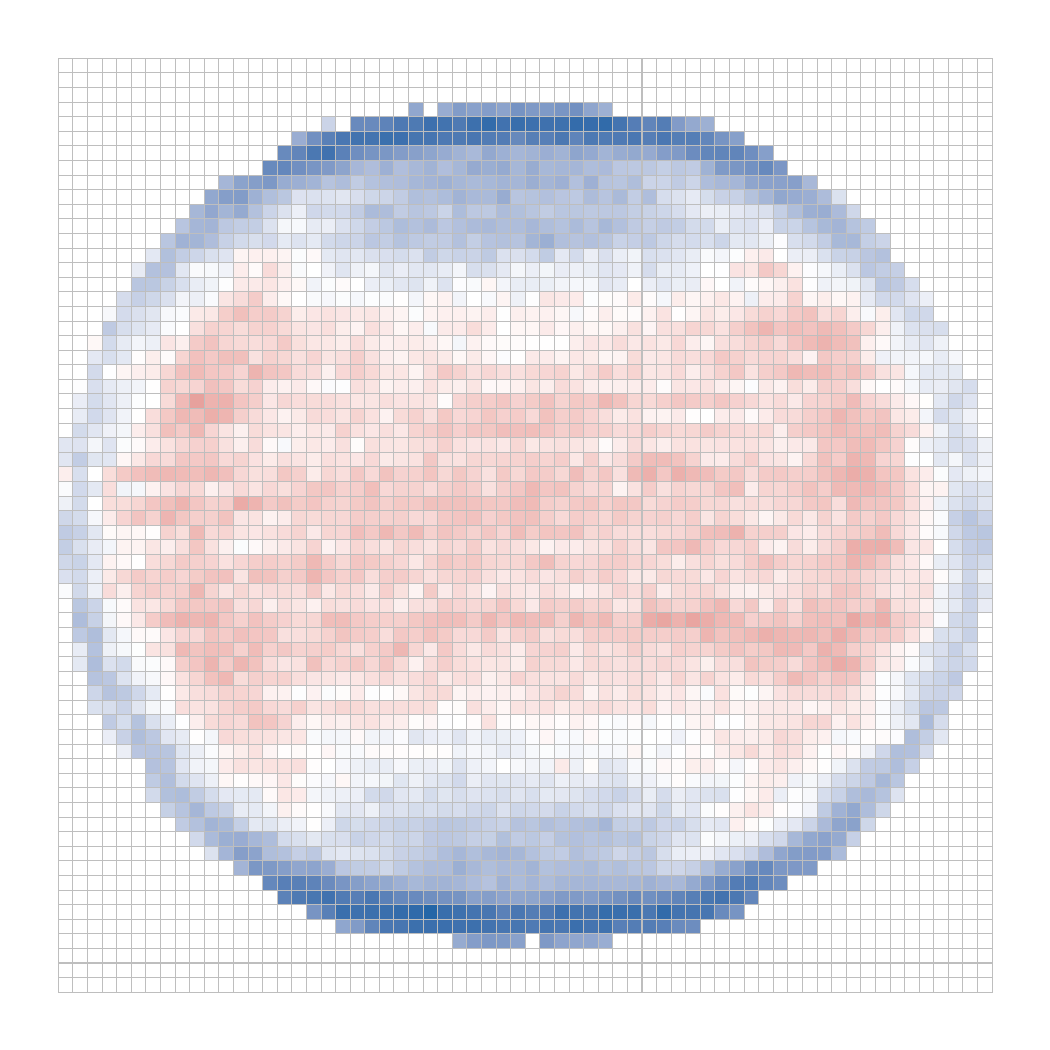}}
\subfloat{\includegraphics[width = 0.33\textwidth]{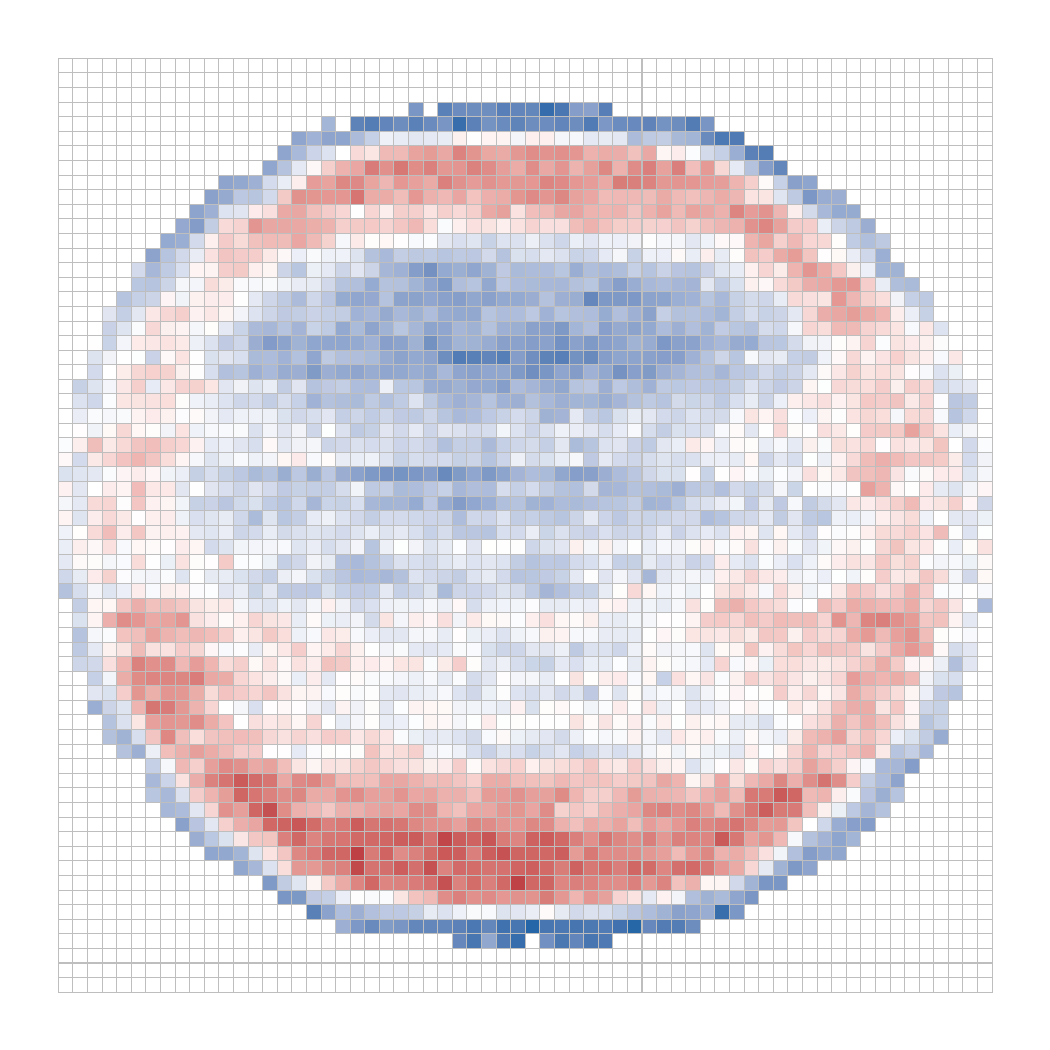}}
\subfloat{\includegraphics[width = 0.33\textwidth]{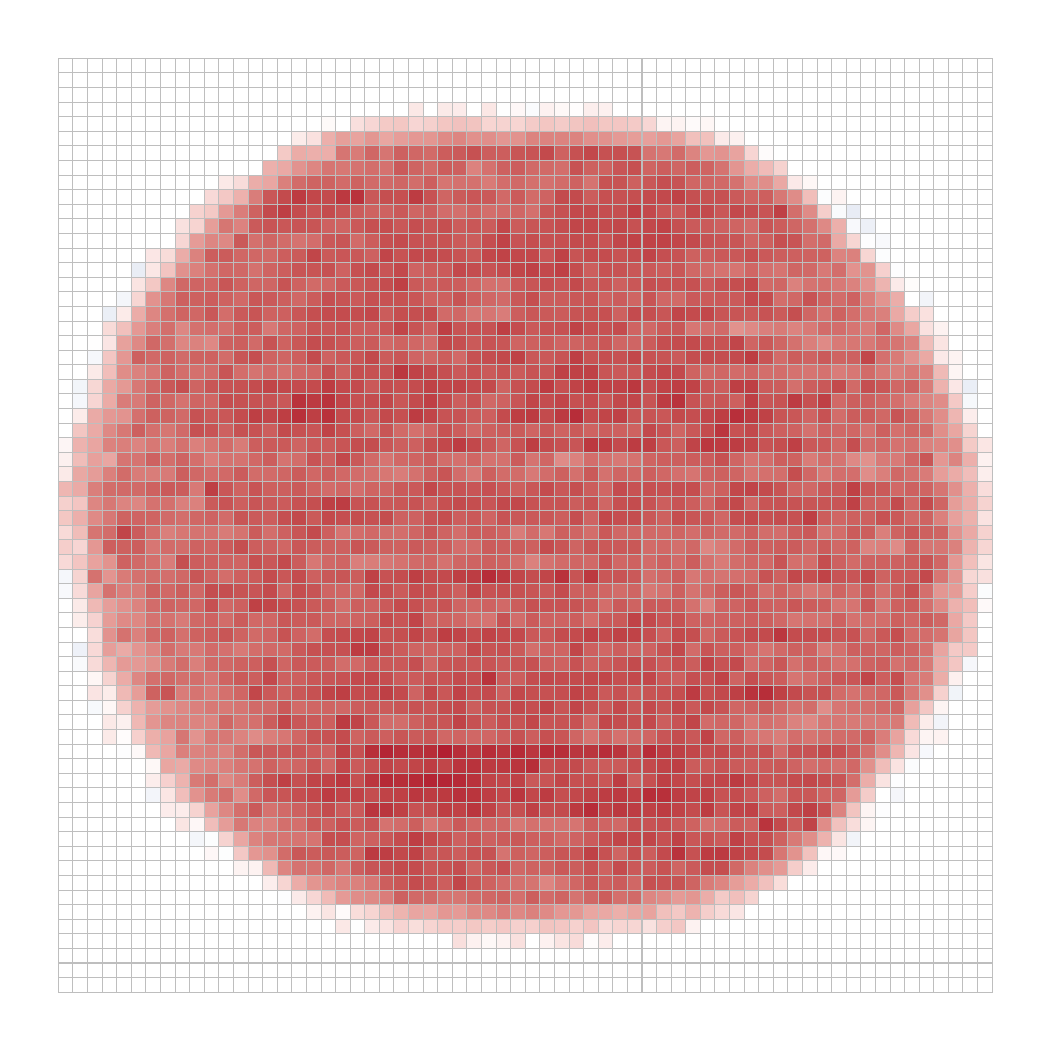}}
\caption{Robust loadings obtained from SCRAMBLE back-transformed to the image space.}
    \label{fig:loadings_wear_robust}
\end{figure}

\begin{figure}
\centering
\subfloat{\includegraphics[width = 0.495\textwidth]{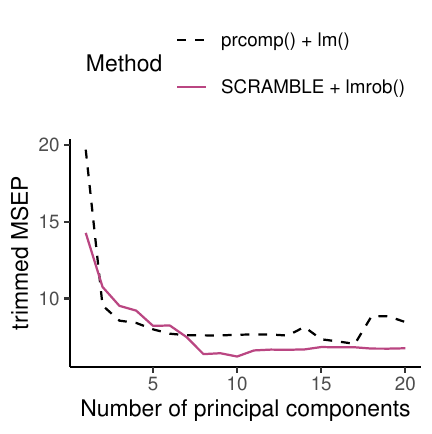}}
\subfloat{\includegraphics[width = 0.495\textwidth]{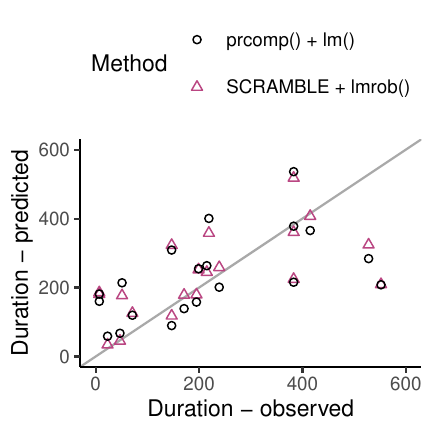}}
\caption{Left: Selection of best number of PCs based on 10\% trimmed MSEP computed on the validation set. Right: Observed vs. predicted alteration duration for the classical and robust approach.}
\label{fig:model_val_test_wear}
\end{figure}

The 10\% trimmed MSEP for the validation set based on different numbers of components is shown on the left side in 
Figure~\ref{fig:model_val_test_wear} for both classical and robust PCR. For classical PCR, the error starts 
at a higher level, possibly indicating that the directions of the first few components are influenced by outliers, and thus they are not as effective for prediction. 
For the optimal number of components, we select that number yielding the smallest trimmed MSEP, resulting in $k_{\text{classical}} = 17$ and $k_{\text{SCRAMBLE}} = 10$ components. Thus, the robust method leads to a smaller number of components, and also to a smaller prediction error.

In the right plot of Figure~\ref{fig:model_val_test_wear}, the observed and predicted values of the alteration duration are shown for both methods for the test set observations. The predictions for higher values of duration for both methods are worse than for values in the beginning or middle of the duration range. In total, the model based on robust PCR performs better, with a trimmed MSEP of $13.76$ for the test set observations, while classical PCR leads to an error of $17.88$ (Figure \ref{fig:model_val_test_wear}).

Using this number of components, we can also reconstruct the data matrices and analyze the reconstruction errors. In 
Figure~\ref{fig:reconstrucion_wear}, the reconstruction errors per variable (pixel) are visualized for both the classical PCA (on the left) and SCRAMBLE (right). While for the classical method, no structure is left in these residuals, it is clearly visible that for the robust method, the border of the wear scars is not reconstructed well. 
This also makes sense because the borders of the balls in the wear scar images are not identical. In fact, the size of the balls in the image can slightly change due to the nature of the experiment, as the balls are placed manually under a microscope, but also due to preprocessing and cutting the images. Obviously, for the robust procedure, this change in size is not relevant for prediction, whereas the classical model takes this variability into account.

\begin{figure}
\centering
\subfloat{\includegraphics[width = 0.495\textwidth]{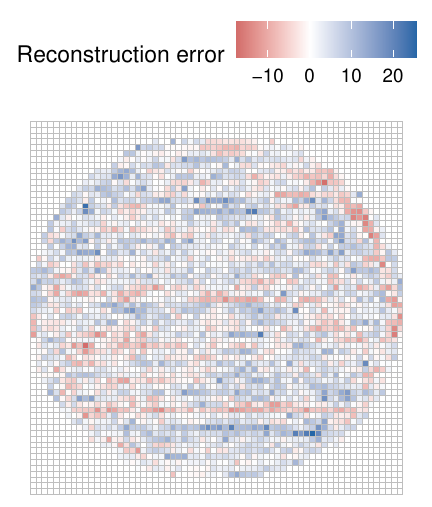}}
\subfloat{\includegraphics[width = 0.495\textwidth]{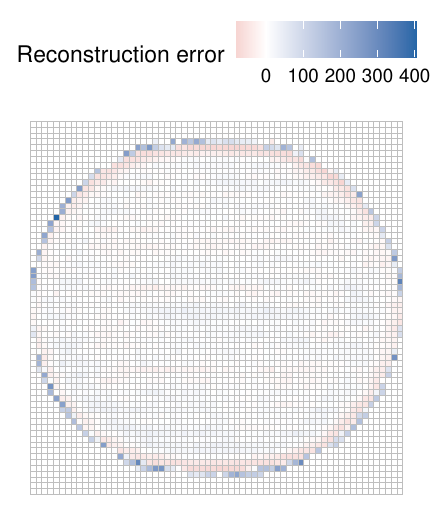}}
\caption{Reconstruction errors per variable, visualized in the image dimensions. Left: PCA via classical SVD, right: PCA via \texttt{SCRAMBLE}. The number of components corresponds to the number leading to the minimum trimmed MSEP for PCR.}
    \label{fig:reconstrucion_wear}
\end{figure}

While the results for classical and robust PCR are not very different, the example still illustrates that the robust method is able to perform better for the majority of the data (in the middle of the duration range), while the classical predictions are influenced by more extreme values. Furthermore, we can use robust diagnostics to get further insight into why the prediction quality between certain values of the response differs. 

\section{Discussion and summary}
\label{sec:sum}

In recent years, cellwise robust methods are becoming increasingly important. This is mainly due to the increased occurrence of high-dimensional data, as a result of modern measurement methods and devices. With high-dimensional data it becomes more likely that an observation contains outliers in single variables, and traditional rowwise (casewise) methods would no longer work if the majority of observations are contaminated. This is also an issue for principal component analysis (PCA), where rowwise robust methods could fail in the presence of many cellwise outliers.

One could think of several different approaches to obtain a cellwise robust PCA method. A first idea could be the identification of cellwise outliers and the replacement of those cells by values which would be expected according to some distributional assumptions~\citep{Rousseeuw2018}. With the cleaned data matrix one could proceed with classical PCA. Even in the casewise robust setting, the approach to detect and correct outlying observations prior to classical PCA would be a way to obtain a casewise robust PCA version. However, outlier detection assumes an underlying model, usually a multivariate normal distribution, and outlier detection/correction identifies/corrects observations or cells according to this model assumption. In cellwise or casewise robust PCA, on the other hand, we are not limited to this distribution. Particularly for PCA based on low-rank approximation, the interest is rather in a robust data reconstruction, where the error loss function utilizes information from the single variables rather than from the joint multivariate distribution, see Equation~\eqref{eq:SVD_RSS_sparse_robust}.

Another approach to cellwise robust PCA is to use a plug-in estimator for the covariance to determine the principal components, for example, the cellwise robust Minimum Covariance Determinant (MCD) estimator~\citep{raymaekers2023cellwise}. This is the cellwise equivalent  to a rowwise robust PCA version where a rowwise robust covariance estimator, such as the MCD~\citep{Rousseeuw1985} is plugged in. While such a procedure is straightforward to implement, it might not be so clear how to include sparsity.

Sparsity, or the natural requirement of interpretability of the principal components is especially important in a high-dimensional setting. For that reason, sparse~\citep{Jolliffe2003} and sparse robust~\citep{Croux2013} PCA versions were introduced which maximize the variance of the components subject to an $L_1$ penalty on the loadings vectors. The gain in sparsity or explainability leads to a loss in explained variance, and this compromise can be formalized by an appropriate objective function~\citep{Croux2013}.

We have introduced a cellwise robust and sparse PCA method using low-rank matrix approximation. The objective 
function can be formulated in a very natural way~\citep{Maronna2008,Croux2013}, and it
combines a robust loss function for the reconstruction error of all
cells of the data matrix with an elastic net penalty on the loadings.
The specific choice of the loss function determines the robustness
properties of the PCA solution~\citep{Maronna2008}.
Both the robust loss function and the incorporation of an $L_1$ or
elastic net penalty leads to computational challenges.
We have developed an algorithm based on manifold learning to optimize the objective function. The $L_1$ penalty was incorporated by the use of a sparsity-inducing penalty, allowing for an approximation by a differentiable function. The choice of appropriate starting values is important, and we compared different approaches. Overall, the algorithm leads to an efficient computation, even for high dimensions $p$ and many observations $n$.
Simulations have demonstrated that the resulting method, called SCRAMBLE (Sparse Cellwise Robust Algorithm for Manifold-based Learning and Estimation),
has superior properties when compared to alternative robust PCA approaches, both in the casewise and cellwise settings. 

We applied the suggested method to two real data examples from tribology and compared the performance with existing estimators, illustrating the usefulness of a cellwise robust and sparse PCA method.

Possible extensions to groupwise PCA or robust data imputation are possible via a modification of the objective function \eqref{eq:loss_function}. 
Furthermore, theoretical robustness properties like the influence function and breakdown point~\citep{Maronna2019} would be interesting topics for future research.

\section*{Funding}
This work was funded by the Austrian COMET-Program (project InTribology2, No. 906860) via the Austrian Research Promotion Agency (FFG) and the federal states of Niederösterreich and Vorarlberg and was carried out at the Austrian Excellence Centre of Tribology (AC2T research GmbH) and the TU Wien. The authors acknowledge TU Wien Bibliothek for financial support through its Open Access Funding Programme.

\end{document}